\documentstyle[11pt]{article}

\newcommand{\K}{\mbox{ , }}
\newcommand{\p}{\mbox{ .}}
\renewcommand{\epsilon}{\varepsilon}
\begin{document}
\title{`Understanding' cosmological bulk viscosity}
\author{Winfried Zimdahl\\ 
Departament de F\'{\i}sica,
Universitat Aut\`{o}noma de Barcelona\\
E-08193 Bellaterra (Barcelona), Spain\\
and\\
Fakult\"{a}t f\"{u}r Physik, Universit\"{a}t Konstanz,
PF 5560 M678\\
D-78434 Konstanz, Germany\thanks{Present address}} 
\date{\today}
\maketitle
\begin{abstract}
A universe consisting of two interacting perfect fluids 
with the same 4-velocity is considered. 
A heuristic mean free time argument is used to show that
the system as a
whole 
cannot be perfect as well but neccessarily
implies a nonvanishing bulk viscosity. 
A new formula for the latter is derived 
and compared with
corresponding results of radiative hydrodynamics.
\end{abstract}
Key words: cosmology: theory - relativity - hydrodynamics -
radiative
transfer\\
\newpage
\section{Introduction}
In the realm of cosmology bulk viscosity is the most favorite 
dissipative phenomenon. Different from shear viscosity and
heat
conductivity, it is compatible with the symmetry
requirements of the
homogeneous and isotropic Friedmann Lema\^{\i}tre
Robertson Walker
(FLRW) universes. In the simplest cosmological models
there is no way
to study entropy producing processes except through
bulk viscosity. 

While this is  obvious on formal grounds it is probably fair to
say that the degree of `understanding' bulk viscosity physically is
by far less than that for shear viscosity or heat flux. 
Although the corresponding effect for a
simple gas is known since the work of Israel (1963) 
and a radiative bulk viscosity coefficient was derived by Weinberg
(1971) 
who also gave an analysis on its r\^{o}le in cosmology, followed by
subsequent investigations of 
Straumann (1976) and Schweizer (1982), there
remained the desire of some more intuitive insight beyond the
involved calculations of radiative hydrodynamics. 
This problem was addressed  in the last part of a paper by 
Udey \& Israel (1982) 
who argued that for a two-fluid universe 
the mechanism responsible
for bulk viscosity is a microscopic heat flux that compensates the 
temperature differences caused by different cooling rates of the two
components. 
Their consideration of this point was based on the following
semiquantitative argument. Let $\tau$ be the characteristic time for
the interaction between both fluids such that during a time
interval $\tau$
the perfect fluid components may be considered as effectively
insulated from each other, resulting in different adiabatic
cooling rates due
to their different equations of state. 

The present paper was inspired by this kind of arguing. 
While we shall confirm below that the difference in the 
cooling rates is indeed the essential point, we shall avoid in our
investigation the introduction of microscopic heat fluxes. 
Although this concept may be helpful on small scales, it seems less
convincing on large scales, e.g., of the Hubble scale. The microscopic
gradients at different points had to conspire in order to produce a
nonvanishing bulk viscosity in homogeneous and isotropic universes,
which, however, is incompatible with the symmetry requirements of the
latter. Moreover, the apparent reduction of a bulk viscous pressure
to heat fluxes is not consistent with the fact that both 
phenomena are basically independent. 

It is the aim of this paper is to show that  {\it different
cooling rates for two perfect fluids are sufficient for the existence
of a nonvanishing bulk viscosity of the system as a whole}. No
additional concept like that of a heat flux over intermolecular
distances has to be used. 
The basic idea is to study a universe of two different interacting 
perfect fluids
and to ask for the conditions under which an effective one-fluid
description is possible. It turns out that this one-fluid universe is
neccessarily dissipative. 

The present paper is semiquantitative throughout. We are not
claiming to improve any of the technically rather complicated
calculations in radiative hydrodynamics. Our objective is to achieve
a kind of phenomenological `understanding' of the bulk pressure
phenomenon in the expanding Universe. 
On this level of description, we shall derive a new formula for the
coefficient of bulk viscosity in a two-fluid system.  

In section 2 the relevant relations for two noninteracting perfect
fluids are
presented and the corresponding cooling rates in an expanding
universe are obtained. 
Section 3 is devoted to an effective one-fluid description for two
fluids with mutual interaction. 
An explicit expression for the coefficient of bulk viscosity is
derived with the help of a mean free time argument. 
The latter result is compared with  work on radiative
hydrodynamics in the context of relativistic kinetic theory in
section 4. Section 5 gives a brief summary of the paper. 
\section{Two-fluid dynamics}
The content of the Universe 
is assumed to be describable by an energy momentum
tensor $T^{ik}$ that is the sum of two different perfect fluid
contributions which share the same 4-velocity:
\begin{equation}
T^{ik} = T^{ik}_{1} + T^{ik}_{2}\K\label{1}
\end{equation}
with ($A$ = 1, 2)
\begin{equation}
T^{ik}_{A} = \rho_{A} u^{i}u^{k} 
+ p_{A} h^{ik}\p \label{2}
\end{equation}
$\rho_{A}$ is the energy density and 
$p_{A}$ is the equilibrium pressure of species $A$.  
$u^{i}$ is the common 4-velocity
and
$h^{ik}$ is the
projection tensor 
$h^{ik} =g^{ik} + u^{i}u^{k}$. 
Let us first deal with the case that the energy momentum 
conservation laws hold for each fluid separately:
\begin{equation}
T^{ik}_{A;k} = 0 \K\label{3}
\end{equation}
implying the energy balances 
\begin{equation}
\dot{\rho}_{A} = - \Theta\left(\rho_{A} + p_{A}\right)\K
\label{4}
\end{equation}
with the fluid expansion 
$\Theta \equiv u^{i}_{;i}$.  $\dot{\rho}_{A} \equiv 
\rho_{A,i}u^{i}$ etc. 
Because of (\ref{3}) both fluids evolve  independently except for
their mutal gravitational coupling.   
The particle flow vector $N_{A}^{i}$ of species $A$ is defined as 
\begin{equation}
N_{A}^{i} = n_{A}u^{i}\K \label{5}
\end{equation}
where $n _{A}$ is the particle number density.
Particle number conservation is expressed by $N_{A;i}^{i} = 0$,
equivalent to 
\begin{equation}
\dot{n}_{A} + \Theta n_{A} = 0 \p\label{6}
\end{equation}

Let us further assume equations of state in the general form
\begin{equation}
p_{A} = p_{A}\left(n_{A}, T_{A}\right) \label{7}
\end{equation}
and
\begin{equation}
\rho_{A} = \rho_{A} \left(n_{A},T_{A}\right)\K \label{8}
\end{equation}
i.e., let 
particle number densities $n_{A}$ and temperatures $T_{A}$ be
our basic
thermodynamical variables. 
The temperatures of both components will be different in general. 

Differentiating  relation (\ref{8}), using the balances 
(\ref{4}) and (\ref{6}) as well as 
the general relation
\begin{equation}
\frac{\partial \rho_{A}}{\partial n_{A}} = 
\frac{\rho_{A} + p_{A}}{n_{A}} 
- \frac{T_{A}}{n_{A}}
\frac{\partial p_{A}}{\partial T_{A}} \K \label{9}
\end{equation}
that follows from the requirement that the entropy is a state
function, 
we find the following expression for the temperature
behaviour:
\begin{equation}
\dot{T}_{A}  = - T_{A}\Theta 
\frac{\partial p_{A}/\partial T_{A}}{\partial \rho_{A}/\partial T_{A}}
\p  \label{10}
\end{equation}
It is obvious that the temperatures of both components behave
differently for different equations of state. 
With $\Theta = 3\dot{R}/R$, where $R$ is the scale factor of the
Robertson-Walker metric, the equations of state 
$p_{1} = n_{1}kT_{1}$, $\rho_{1} = 3n_{1}kT_{1}$ reproduce the well
known $T_{1} \sim R^{-1}$ behaviour for radiation. 
With  $p_{2} = n_{2}kT_{2}$, $\rho_{2} = n_{2}mc^{2} +
\frac{3}{2}n_{2}kT_{2}$ one obtains $T_{2} \sim R^{-2}$ 
for matter. 
\section{Effective one-fluid dynamics}
The hitherto independent fluids are now allowed to interact. We try
to find an effective one fluid description for the Universe as a
whole, characterized by the particle number density $n = n_{1} +
n_{2}$ and an equilibrium temperature $T$. 
The overall equations of state are 
\begin{equation}
p = p\left(n, T\right) \label{11}
\end{equation}
and
\begin{equation}
\rho = \rho \left(n,T\right)\K \label{12}
\end{equation}
where $p$ is the equilibrium pressure and $\rho$ is the energy
density of the system as a whole. 
The equilibrium temperature $T$ is {\it defined} by 
(cf. Udey \& Israel 1982) 
\begin{equation}
\rho_{1}\left(n_{1},T_{1}\right) + \rho_{2}\left(n_{2},T_{2}\right) 
= \rho \left(n,T\right)
\p \label{13}
\end{equation}
As we shall show below this implies
\begin{equation}
p_{1}\left(n_{1},T_{1}\right) + p_{2}\left(n_{2},T_{2}\right) 
\neq p\left(n,T\right)
\p \label{14}
\end{equation}
For perfect fluids 
the difference between both sides of the latter inequality is the
viscous pressure $\pi$
\footnote{For real fluids this is not neccessarily true 
(see, e.g. Kirkwood \& Oppenheim 1961 chapter 7.7)}:
\begin{equation}
\pi = 
p_{1}\left(n_{1},T_{1}\right) + p_{2}\left(n_{2},T_{2}\right) 
- p\left(n,T\right)
\p \label{15}
\end{equation}
The existence of a nonvanishing viscous pressure is a consequence of
the different temperature evolution laws of the subsystems. 
This is most easily understood by the following simple mean free time
argument. 
Let $\tau$ be the characteristic mean free time for the interaction
between both components. 
The time $\tau$ is assumed to be much larger than the characteristic
interaction times within each of the components. Consequently, the
latter may be regarded as perfect fluids on time scales of the order of
$\tau$. The interaction between the fluids is modelled by
`collisional' events, where $\tau$ plays the r\^{o}le of a mean free
time between these `collisions'. During the time interval $\tau$,
i.e., between subsequent interfluid interaction events, both
components then evolve according their internal perfect fluid
dynamics, given by (\ref{4}), (\ref{6}) and (\ref{10}).  
Assume that through this interaction an element of the cosmic fluid
is in equilibrium at a proper time $\eta_{0}$ at a temperature 
$T\left(\eta_{0}\right) = T_{1}\left(\eta_{0}\right) 
= T_{2}\left(\eta_{0}\right)$ with 
$p\left(\eta_{0}\right) = p_{1}\left(\eta_{0}\right) + 
p_{2}\left(\eta_{0}\right)$. 
Here,  
$p\left(\eta_{0}\right)$ and $p_{A}\left(\eta_{0}\right)$  
are shorts for 
$p\left[n\left(\eta_{0}\right),T\left(\eta_{0}\right)\right]$ 
and 
$p_{A}\left[n_{A}\left(\eta_{0}\right),
T_{A}\left(\eta_{0}\right)\right]$, respectively. 
Using the condition (\ref{13}) 
at the proper time 
$\eta_{0} + \tau$ up to first order in $\tau$, i.e., 
\begin{equation}
\rho_{A}\left(\eta_{0} + \tau\right) = 
\rho_{A}\left(\eta_{0}\right) + \tau
\dot{\rho}_{A}\left(\eta_{0}\right) + ...\label{16}
\end{equation} 
and 
\begin{equation}
\rho\left(\eta_{0} + \tau\right) = 
\rho\left(\eta_{0}\right) + \tau
\dot{\rho}\left(\eta_{0}\right) + ... \label{17}
\end{equation}
where 
$\rho_{A}\left(\eta_{0}\right) \equiv 
\rho_{A}\left[n_{A}\left(\eta_{0}\right),
T_{A}\left(\eta_{0}\right)\right]$ 
and 
$\rho\left(\eta_{0}\right) \equiv 
\rho\left[n\left(\eta_{0}\right),
T\left(\eta_{0}\right)\right]$, 
applying (\ref{4})  
on the l.h.s. of (\ref{13})  and the general relation 
\begin{equation}
\frac{\partial \rho}{\partial n} = 
\frac{\rho + p}{n} 
- \frac{T}{n}\frac{\partial p}{\partial T} \K \label{18}
\end{equation}
on its r.h.s., one finds 
\begin{equation}
\dot{T}\left(\eta_{0}\right)  = - T
\frac{\partial p/\partial T}{\partial \rho /\partial T}
\Theta   \label{19}
\end{equation}
at the point $\eta = \eta_{0}$. 

During the following time interval $\tau$, i.e., until a subsequent
`collision', the subsystems move freely according to their proper
dynamics given by (\ref{4}), (\ref{6}) and (\ref{10}).   
At the time $\eta_{0} + \tau$ we have 
$T\left(\eta_{0} + \tau\right) \neq T_{1}\left(\eta_{0} + \tau\right) 
\neq T_{2}\left(\eta_{0} + \tau\right) 
\neq T\left(\eta_{0} + \tau\right)$  
in general. 
With 
\begin{equation}
T_{A}\left(\eta_{0} + \tau\right) = 
T_{A}\left(\eta_{0}\right) + \tau
\dot{T}_{A}\left(\eta_{0}\right) + ... \label{20}
\end{equation}
and 
\begin{equation}
T\left(\eta_{0} + \tau\right) = 
T\left(\eta_{0}\right) + \tau
\dot{T}\left(\eta_{0}\right) + ... \label{21}
\end{equation}
we find from (\ref{10}) and (\ref{19}) 
\begin{equation}
T_{1} - T_{2}  = - \tau \Theta T\left(
\frac{\partial p_{1}/\partial T}{\partial \rho_{1}/\partial T} - 
\frac{\partial p_{2}/\partial T}{\partial \rho_{2}/\partial T}\right) 
 \K  \label{22}
\end{equation}
\begin{equation}
T_{1} - T  = - \tau \Theta T\left(
\frac{\partial p_{1}/\partial T}{\partial \rho_{1}/\partial T} - 
\frac{\partial p/\partial T}{\partial \rho/\partial T}\right) 
 \K  \label{23}
\end{equation}
\begin{equation}
T_{2} - T  = - \tau \Theta T\left(
\frac{\partial p_{2}/\partial T}{\partial \rho_{2}/\partial T} - 
\frac{\partial p/\partial T}{\partial \rho/\partial T}\right) 
 \K  \label{24}
\end{equation}
up to first order in $\tau$. 
Due to the different cooling rates (\ref{10}) and (\ref{19})  there
occur temperature differences at any point of the expanding fluid:
Both differences in the temperatures of the components, i.e., between
$T_{1}$ and $T_{2}$, and differences between the temperature of each
of the components and the temperature $T$ of the system as a whole. 

In order to arrive at our conclusion $\pi \neq 0$ one has simply to
consider the sum of the partial pressures $p_{1}$ and $p_{2}$ at 
$\eta = \eta_{0} + \tau$ up to first order in $\tau$. 
The latter may be written as 
\begin{eqnarray}
p_{1}\left(n_{1},T_{1}\right) + p_{2}\left(n_{2},T_{2}\right) 
&=& p_{1}\left(n_{1},T\right) + p_{2}\left(n_{2},T\right) 
\nonumber\\ &&
+ \left(T_{1} - T\right)\frac{\partial p_{1}}{\partial T} 
+ \left(T_{2} - T\right)\frac{\partial p_{2}}{\partial T} 
\p \label{25}
\end{eqnarray}
Inserting the temperature differences $T_{1} - T$ and $T_{2} - T$ at
$\eta_{0} + \tau$ from (\ref{23}) and (\ref{24}), we find
\begin{eqnarray}
p_{1}\left(n_{1},T_{1}\right) + p_{2}\left(n_{2},T_{2}\right) 
&=& p\left(n,T\right) 
- \tau T\Theta \left[ 
\frac{\partial p_{1}}{\partial T} 
\left(
\frac{\partial p_{1}/\partial T}{\partial \rho_{1}/\partial T} - 
\frac{\partial p/\partial T}{\partial \rho/\partial T}\right) 
\right. \nonumber\\
&&\left. \mbox{\ \ \ \ \ \ }
+ \frac{\partial p_{2}}{\partial T} 
\left(
\frac{\partial p_{2}/\partial T}{\partial \rho_{2}/\partial T} - 
\frac{\partial p/\partial T}{\partial \rho/\partial T}\right)\right]
\K \label{26}
\end{eqnarray}
where $p\left(n, T\right) \equiv p_{1}\left(n_{1}, T\right) 
+ p_{2}\left(n_{2}, T\right)$ was used. 
Applying the zeroth-order relations 
$\partial p/\partial T = \partial p_{1}/\partial T 
+ \partial p_{2}/\partial T$ and 
$\partial \rho/\partial T = \partial \rho_{1}/\partial T 
+ \partial \rho_{2}/\partial T$ in 
the bracket on the r.h.s. of (\ref{26}) one gets, 
after a simple rearrangement, the first-order result
\begin{equation}
p_{1}\left(n_{1},T_{1}\right) + p_{2}\left(n_{2},T_{2}\right) 
= p\left(n,T\right) + \pi
\K \label{27}
\end{equation}
where 
\begin{equation}
\pi \equiv \tau T\Theta \frac{\partial \rho}{\partial T} 
\left(
\frac{\partial p_{1}/\partial T}{\partial \rho_{1}/\partial T} - 
\frac{\partial p/\partial T}{\partial \rho/\partial T}\right) 
\left(
\frac{\partial p_{2}/\partial T}{\partial \rho_{2}/\partial T} - 
\frac{\partial p/\partial T}{\partial \rho/\partial T}\right) 
\label{28}
\end{equation}
is generally different from zero. 
This proves our initial statement that a system of two interacting 
perfect fluids
is not perfect as well. While the energy momentum tensors of the
subsystems are given by (\ref{2}), the system as a whole is
characterized
by 
\begin{equation}
T^{ik} = \rho u^{i}u^{k} 
+ \left(p + \pi\right) h^{ik}\p \label{29}
\end{equation}
To separate bulk viscosity from any other dissipative phenomenon we
have ignored here the possibility of nonvanishing heat fluxes and
shear stresses in inhomogeneous and anisotropic cosmological models. 

From the definition
\begin{equation}
\pi = - \zeta \Theta \K \label{30}
\end{equation}
of the bulk viscosity $\zeta$, the latter is found to be given by
\begin{equation}
\zeta = - \tau T \frac{\partial \rho}{\partial T} 
\left(
\frac{\partial p_{1}}{\partial \rho_{1}} - 
\frac{\partial p}{\partial \rho}\right) 
\left(
\frac{\partial p_{2}}{\partial \rho_{2}} - 
\frac{\partial p}{\partial \rho}\right) 
\K\label{31}
\end{equation}
where 
\begin{equation}
\frac{\partial p_{A}}{\partial \rho_{A}} \equiv  
\left(\frac{\partial p_{A}}{\partial \rho_{A}}\right)_{n_{A}} \equiv 
\frac{\partial p_{A}/\partial T}{\partial \rho_{A}/\partial T} 
\K 
\frac{\partial p}{\partial \rho} \equiv  
\left(\frac{\partial p}{\partial \rho}\right)_{n} \equiv 
\frac{\partial p/\partial T}{\partial \rho/\partial T} 
\p\label{32}
\end{equation}
For `ordinary' matter $\partial p_{A}/\partial \rho_{A}$ lies in the
range $1/3 \leq \partial p_{A}/\partial \rho_{A} \leq 2/3$ (see the
equations of state below (\ref{10})). The lower limit corresponds to
radiation, the upper one to matter. 
$\partial p/\partial \rho$ will take a
value intermediate between 
$\partial p_{1}/\partial \rho_{1}$ and 
$\partial p_{2}/\partial \rho_{2}$. 
Assuming without loss of generality 
$\partial p/\partial \rho > 
\partial p_{1}/\partial \rho_{1}$, we
shall have 
$\partial p/\partial \rho < 
\partial p_{2}/\partial \rho_{2}$. 
Consequently, $\zeta \geq 0$, i.e., the entropy production is 
positive which agrees with the second law of thermodynamics 
(see, e.g., de Groot, van Leeuwen \& van Weert 1980). 

To the best of our knowledge the expression (\ref{31}) for the
coefficient of bulk viscosity is a new result . Although based on
heuristic arguments its structure is rather general. 
Formula (\ref{31}) is valid for general equations of state 
(\ref{7}), (\ref{8}) and (\ref{11}), (\ref{12}). Until now we did not
specify to the case that one of the components obeys the equations of
state for radiation. 
\section{Relation to radiative hydrodynamics}
The expression (\ref{31}) for the coefficient of bulk viscosity is
similar
but not identical to the corresponding expressions of radiative
hydrodynamics 
found by Weinberg (1971), Straumann (1976), 
Schweizer (1982), Udey \& Israel (1982) and 
Pav\'{o}n, Jou \& Casas-V\'{a}zquez (1983).  
The reason for this is a physical one and not due to the
semiquantitative nature of our considerations. 
Both components of the system are treated as fluids with different
equations of state in our setting. 
The final result (\ref{31}) is therefore 
symmetric under a change of the labels $1$
and $2$ that identify both components. 
In the work of Weinberg (1971), Straumann (1976), 
Schweizer (1982), Udey \& Israel (1982) and 
Pav\'{o}n, Jou \& Casas-V\'{a}zquez (1983)  
on the other hand, both components are treated asymmetrically. 
While one of the
components is a fluid as well, the second one, a radiation
component, is described with the help of kinetic theory. 
The main asymmetry lies in the assumption of 
the mentioned authors that the
radiation component is allowed to deviate from local equilibrium
while the fluid component is not. Of course, the result for the
coefficient of bulk viscosity is not symmetric in the components
either and a coincidence with (\ref{31}) cannot be expected. 

It might be useful to compare some of the basic relations of
radiative hydrodynamics 
with our framework. For definitenes, let component $1$ of our analysis
be the radiation and component $2$ the material component. 
Since there does not appear a separate radiation temperature in the
papers by 
Weinberg (1971), Straumann (1976), 
Schweizer (1982), Udey \& Israel (1982) and 
Pav\'{o}n, Jou \& Casas-V\'{a}zquez (1983), 
let us eliminate the latter in (\ref{13}). 
With $\rho_{1}\left(n_{1},T_{1}\right) = 
\rho_{1}\left(n_{1},T_{2}\right) + \left(T_{1} - T_{2}\right)
\partial \rho_{1}/\partial T + ...$ we find, up to first order in the
temperature difference, 
\begin{equation}
\rho \left(n,T\right) = 
\rho_{1}\left(n_{1},T_{2}\right) + \rho_{2}\left(n_{2},T_{2}\right) 
+ \hat{\rho}
\K \label{33}
\end{equation}
where 
\begin{equation}
\hat{\rho} \equiv  \left(T_{1} - T_{2}\right) 
\frac{\partial \rho_{1}}{\partial T}
\p \label{34}
\end{equation}
Using (\ref{22}) with (\ref{32})  we have 
\begin{equation}
\hat{\rho} = - \tau T \Theta\frac{\partial \rho_{1}}{\partial T} 
\left(
\frac{\partial p_{1}}{\partial \rho_{1}} - 
\frac{\partial p_{2}}{\partial \rho_{2}}\right) 
\p\label{35}
\end{equation}
For comparison with the results of Udey \& Israel (1982) for radiative
hydrodynamics it is helpful to rewrite the latter expression as 
\begin{equation}
\hat{\rho} = \hat{\rho}_{UI} +  
\tau T \Theta\frac{\partial \rho_{1}}{\partial T} 
\left(
\frac{\partial p_{2}}{\partial \rho_{2}} - 
\frac{\partial p}{\partial \rho}\right) 
\K\label{36}
\end{equation} 
where
\begin{equation}
\hat{\rho}_{UI} = - \tau T \Theta\frac{\partial \rho_{1}}{\partial T} 
\left(
\frac{\partial p_{1}}{\partial \rho_{1}} - 
\frac{\partial p}{\partial \rho}\right) 
\p\label{37}
\end{equation}
With (\ref{36}) and (\ref{37}) 
relation (\ref{33}) may be compared with formula 
(31) of Udey \& Israel (1982), which, in our notation, reads 
\begin{equation}
\rho \left(n,T\right) = 
\rho\left(n,T_{2}\right) + aT^{4}B \p \label{38}
\end{equation}
The quantity $\hat{\rho}$ in our equation (\ref{33}) 
is the counterpart of the term $aT^{4}B$ in   
Udey \& Israel (1982), which describes the
deviation of the radiation component from equilibrium. 
With $T\partial
\rho_{1}/\partial T = 4aT^{4}$ and 
$\partial p_{1}/\partial \rho_{1} = 1/3$ for radiation and
specification of (35b), (45) and (47) in 
Udey \& Israel (1982) to the Eckart case
(i.e., neglecting relaxation and cross effects), we find that 
$aT^{4}B$ in (\ref{38}) (equation (31) in Udey \& Israel 1982) 
coincides with $\hat{\rho}_{UI}$ 
of our equation (\ref{37}). 
The circumstance that there exists a difference between 
$\hat{\rho}$ and $\hat{\rho}_{UI}$ reflects the above mentioned fact
that, different from radiative hydrodynamics, in our setting both
components are allowed to deviate from equilibrium. 
For $\partial p_{2}/\partial \rho_{2} \approx 
\partial p/\partial \rho$ this difference becomes negligible and we
have $\hat{\rho} \approx \hat{\rho}_{UI}$. 

A similar statement holds for the difference of the temperatures
$T_{2} - T$. 
By virtue of the zeroth-order identity
\begin{equation}
\frac{\partial \rho}{\partial T} 
\left(
\frac{\partial p}{\partial \rho} - 
\frac{\partial p_{2}}{\partial \rho_{2}}\right)
= 
\frac{\partial \rho_{1}}{\partial T} 
\left(
\frac{\partial p_{1}}{\partial \rho_{1}} - 
\frac{\partial p_{2}}{\partial \rho_{2}}\right) 
\label{39}
\end{equation}
the temperature difference (\ref{24}) may be written as
\begin{eqnarray}
T - T_{2} &=& - \tau T \Theta\frac{\partial \rho_{1}/\partial T}
{\partial \rho/\partial T} 
\left(
\frac{\partial p_{1}}{\partial \rho_{1}} - 
\frac{\partial p_{2}}{\partial \rho_{2}}\right) \nonumber\\
&=& \left(T - T_{2}\right)_{W} 
+ \tau T \Theta\frac{\partial \rho_{1}/\partial T}
{\partial \rho/\partial T} 
\left(
\frac{\partial p_{2}}{\partial \rho_{2}} - 
\frac{\partial p}{\partial \rho}\right)
\K\label{40}
\end{eqnarray}
where 
\begin{equation}
\left(T - T_{2}\right)_{W} =  
- \tau T \Theta\frac{\partial \rho_{1}/\partial T}
{\partial \rho/\partial T} 
\left(
\frac{\partial p_{1}}{\partial \rho_{1}} - 
\frac{\partial p}{\partial \rho}\right)
\K\label{41}
\end{equation}
if specified to radiation, coincides with 
Weinberg's  relation (2.38) (Weinberg 1971) 
for radiative hydrodynamics. 
The condition for $T - T_{2} \approx  \left(T - T_{2}\right)_{W}$ 
is again $\partial p_{2}/\partial \rho_{2} \approx 
\partial p/\partial \rho$. 

Using the corresponding intermediate steps that led to (\ref{33}) 
now for the sum of the
pressures, we get the first-order relation 
\begin{equation}
p_{1}\left(n_{1},T_{1}\right) + p_{2}\left(n_{2},T_{2}\right)  = 
p_{1}\left(n_{1},T_{2}\right) + p_{2}\left(n_{2},T_{2}\right) 
+ \hat{p}
\K \label{42}
\end{equation}
with 
\begin{equation}
\hat{p} \equiv  \left(T_{1} - T_{2}\right) 
\frac{\partial p_{1}}{\partial T} = 
\frac{\partial p_{1}}{\partial \rho_{1}}
\hat{\rho}
\p \label{43}
\end{equation}
For the first two terms on the r.h.s. of (\ref{42})  we may write,
again up to first order, 
\begin{equation}
p_{1}\left(n_{1},T_{2}\right) + p_{2}\left(n_{2},T_{2}\right) 
\equiv p\left(n,T_{2}\right) = p\left(n,T\right) + 
\left(T_{2} - T\right) 
\frac{\partial p}{\partial T}
\p \label{44}
\end{equation}
From (\ref{40})  and (\ref{35}) one has 
\begin{equation}
T_{2} - T = - \left(\frac{\partial \rho}{\partial T}\right)^{-1}
\hat{\rho}
\K \label{45}
\end{equation}
and, consequently, 
\begin{equation}
p_{1}\left(n_{1},T_{1}\right) + p_{2}\left(n_{2},T_{2}\right) 
= p\left(n,T\right) + \pi 
\K \label{46}
\end{equation}
with 
\begin{equation}
\pi =  
\left(
\frac{\partial p_{1}}{\partial \rho_{1}} - 
\frac{\partial p}{\partial \rho}\right) 
\hat{\rho}
\p\label{47}
\end{equation}
This is relation (33) of Udey \& Israel (1982). 
 
In order to avoid misunderstandings we point out that it was not
neccessary for our derivation of formula (\ref{31}) 
to introduce the quantities $\hat{\rho}$ 
and $\hat{p}$. These quantities are useful, however, for the
comparison with work done in radiative hydrodynamics. 

By virtue of the identity (\ref{39}), $\zeta$ from (\ref{31})  
may be written as
\begin{equation}
\zeta =  \tau T \frac{\partial \rho_{1}}{\partial T} 
\left(
\frac{\partial p_{1}}{\partial \rho_{1}} - 
\frac{\partial p_{2}}{\partial \rho_{2}}\right) 
\left(
\frac{\partial p_{1}}{\partial \rho_{1}} - 
\frac{\partial p}{\partial \rho}\right) 
\p\label{48}
\end{equation}
While it is generally not to be expected that this expression
coincides with Weinberg's coefficient (Weinberg 1971) 
\begin{equation}
\zeta_{W} =  \tau T \frac{\partial \rho_{1}}{\partial T} 
\left(
\frac{\partial p_{1}}{\partial \rho_{1}} - 
\frac{\partial p}{\partial \rho}\right)^{2}  
\K\label{49}
\end{equation}
it is obvious that both results 
are the closer  the
better the approximation of 
$\partial p/\partial \rho$ 
by 
$\partial p_{2}/\partial \rho_{2}$ in (\ref{31}) will be. 
In other words, the results (\ref{48}) and (\ref{49}) are similar if
the matter component dominates the behaviour of the system as a
whole. 

The bulk viscosity coefficients $\zeta$ and $\zeta_{W}$ are related by
\begin{equation}
\zeta = \zeta_{W} +  \tau T \frac{\partial \rho_{1}}{\partial T} 
\left(
\frac{\partial p}{\partial \rho} - 
\frac{\partial p_{2}}{\partial \rho_{2}}\right) 
\left(
\frac{\partial p_{1}}{\partial \rho_{1}} - 
\frac{\partial p}{\partial \rho}\right) 
\p\label{50}
\end{equation}
Using  (\ref{31}) on the r.h.s of the latter equation we obtain the
following relation between $\zeta$ and $\zeta_{W}$:
\begin{equation}
\frac{\partial \rho_{2}}
{\partial \rho} \zeta = \zeta_{W}
\K\label{51}
\end{equation}
where
\begin{equation}
\frac{\partial \rho_{2}}{\partial \rho} \equiv 
\frac{\partial \rho_{2}/\partial T}
{\partial \rho/\partial T} 
\p\label{52}
\end{equation}
Relation (\ref{51}) shows again that the generally different
expressions for $\zeta$ and $\zeta_{W}$ will become similar if the
overall behaviour of the system is determined by the fluid component
$2$. 

For the specific case of a mixture of radiation and matter with
equations of state 
$p_{1} = n_{1}kT_{1}$, $\rho_{1} = 3n_{1}kT_{1}$,  $p_{2} =  
n_{2}kT_{2}$, 
$\rho_{2} = n_{2}mc^{2} +
\frac{3}{2}n_{2}kT_{2}$ the bulk viscosity coefficient 
(\ref{31})  reduces to 
\begin{equation}
\zeta = \frac{\tau}{3}n_{1}kT  \frac{n_{2}}{2n_{1} + n_{2}} 
\p\label{53}
\end{equation}
Since $\partial \rho_{2}/\partial \rho = 
n_{2}/\left(2n_{1} + n_{2}\right)$ one has $\zeta \approx \zeta_{W}$
for $n_{1} \ll n_{2}$. 
The lower the ratio $n_{1}/n_{2}$ of the photon number density to the
number density of the matter particles the closer to unity is the
ratio $\zeta_{W}/\zeta$. 

A fluid description of the Universe makes sense as long as $\tau \ll
H^{-1}$, where $H \equiv \Theta/3$ is the Hubble parameter. 
Combining (\ref{53}) with (\ref{30}) we find that this condition is
consistent with $\mid\pi\mid \ll p_{A}$, i.e., the magnitude of the
nonequilibrium part $\pi$ of the pressure is much smaller than the
equilibrium pressures as it is neccessary for a first-order approach
like that of the present paper to be valid. 
\section{Summary}
This paper is a heuristic attempt to clarify the origin 
of bulk viscosity in the expanding universe. 
Characterizing the interaction between two different fluids, each of
them perfect on its own, by an effective mean free time parameter
$\tau$ and assuming a free evolution of both components according to
their internal perfect fluid dynamics during the time interval
$\tau$, i.e., between subsequent 
interfluid interaction events, the different
cooling rates (due to different equations of state) of the components
lead to a nonvanishing bulk pressure of the system as a
whole. 
A new formula for the coefficient of bulk viscosity of a two-fluid
mixture was obtained and the relation of this expression to the
results of radiative hydrodynamics was clarified. \\

\ \\
{\bf Acknowledgement} \\
The warm hospitality of the Grup de F\'{\i}sica Estad\'{\i}stica of
the Autonomous University of Barcelona 
and useful discussions with  D. Pav\'{o}n and D. Jou, who critically
read the manuscript, 
J. Casas-V\'{a}zquez as well as with R. Maartens, Portsmouth, 
 are gratefully acknowledged.

\ \\
{\bf  REFERENCES}
\ \\
de Groot S.R., van Leeuwen W.A., van Weert Ch.G., 1980,\\ 
$\mbox{\ \ \ \ \ }$Relativistic
Kinetic Theory. North-Holland, Amsterdam\\
Israel W., 1963, J. Math. Phys., 4, 1163\\
Kirkwood J.G., Oppenheim I., 1961,\\ 
$\mbox{\ \ \ \ \ }$Chemical Thermodynamics. 
McGraw-Hill, New York\\
Pav\'{o}n D., Jou D., Casas-V\'{a}zquez J., 1983, 
J. Phys.A, 16, 775\\
Schweizer M. A., 1982, Astrophys. J., 258, 798\\
Straumann N., 1976, Helv. Phys. Acta, 49, 269\\
Udey N., Israel W., 1982, Mon. Not. R. Astr. Soc., 199, 1137\\
Weinberg S., 1971, Astrophys. J., 168, 175  

\end{document}